\documentclass{acm_proc_article-sp}
\toappear{F.Gargiulo, S.Huet: When group level is different from the population level: an adaptive network with the Deffuant model. In Frank Schweitzer, Akira Namatame,
 Hideyuki Nakashima, and Satoshi Kurihara (eds.): Proc.\ of 5th
 Int.\ Workshop on Emergent Intelligence on Networked Agents
 (WEIN 2010) at AAMAS 2010; May, 10, 2010, Toronto, Canada.}
\begin{document}

\title{When group level is different from the population level: an adaptive network with the Deffuant model}

\numberofauthors{2}

\author{
% 1st. author
\alignauthor
Floriana Gargiulo\\
       \affaddr{LISC,Cemagref}\\
       \affaddr{24 Avenue de Landais}\\
       \affaddr{63170 AUBIERE, France}\\
       \email{floriana.gargiulo@cemagref.fr}
% 2nd. author
\alignauthor
Sylvie Huet\\
       \affaddr{LISC,Cemagref}\\
       \affaddr{24 Avenue de Landais}\\
       \affaddr{63170 AUBIERE, France}\\
       \email{sylvie.huet@cemagref.fr}
}

\maketitle
\begin{abstract}
We propose a model coupling the classical opinion dynamics of the bounded confidence model, proposed by Deffuant et al., with an adaptive network forming community or group structure. At each step, an individual can decide if it changes groups or interact on its opinion with one of its internal or external neighbour. If it decides to look at the group level, it changes group if its opinion is far from the average of its group from more than a threshold. If it is the case, it joins the group which has proportionally the closest average opinion from its. If it decides to interact with one of its neighbour, it becomes closer in opinion to it when its opinion and the one of the selected-to-interact neighbour are less distant from the threshold.\\
From the study of this coupled model, we discover some surprising behaviours compared to the known behaviour of the Deffuant bounded confidence model(BC): The coupled model exhibits a total consensus for an threshold value lower than the BC model; the distribution of sizes of the groups changes: some groups become larger while other decrease in size, sometimes until containing only one individual. From the point of view of the groups, the consensus remains for a large set of threshold values while, looking at the population level, there are a lot of opinion clusters. 
\end{abstract}

\section{Introduction}

Other's opinion is a source of cognitive inconsistency! That is what Festinger \cite{festinger1957Dissonance} argued adding that it is experienced as dissonance. According to him, the dissonance is a psychological discomfort or an aversive drive state that people are motivated to reduce, just as they are motivated to reduce hunger. In his balance theory, \cite{heider1946} used a similar concept and called it imbalance. More recently, \cite{MatzWood2005} showed that, as the dissonance and balance theories suggest, the disagreement from others in a group produces cognitive inconsistency and the negative states of dissonance or imbalance.\\

The groups are a privileged place of interaction between people and the exchange with others can lead to dissonance. They are thus at the same time they entity creating dissonance and the one reducing it. Indeed, three strategies can be chosen to reduce its dissonance created by the heterogeneity of the opinion inside its group: changing one's own opinion to agree with others in the group, influencing others to change their opinions, or joining a different, attitudinally more congenial group. The three ones reduce dissonance \cite{MatzWood2005}. \\

The two first relates to the individual interactions which are often based on similarity and have been extensively studied in the attraction paradigm \cite{Byrne1971} and other theories on interpersonal interactions as the social judgment theory \cite{SherifHovland1961}. The third can be linked to the personal external network of the individual. Indeed, its external to its own group neighbour give it some information about the characteristics of their group and can introduce it.\\

Starting from these strategies, the present paper studies a simulation model reproducing their main aspects in order to better understand the link between the individual choices and the organization of the society into groups. We will model the group concept as the one of community is: based on a social network where an individual has most of its links to its own group and a minor part to the other groups. The interaction process between individuals will be modeled by one of the classical opinion dynamic model.\\

On the one hand, the social network, with or without communities, coupled to various reaction processes, has been intensively studied in the last decades \cite{pastor2004evolution}: from epidemics \cite{pastor2001epidemic}, \cite{pastor2002immunization}, to malware diffusion in electronic technology \cite{hu2009wifi}, collective behaviors \cite{helbing2000simulating}, innovation diffusion and opinion dynamics \cite{castellano2009statistical}, \cite{galam2008sociophysics}. \\

On the other hand, many opinion dynamic models have been proposed to study the spreading of opinion: some of these models describe opinion as a discrete Boolean choice, like the Voter model \cite{clifford1973model}, \cite{holley1978survival} or the Sznajd model \cite{stauffer2002sociophysics}. These formalizations can describe, for example, the positions on elections in majoritarian systems (where only two parties are present). Other models take into account the fact that, for some kinds of situation, people can have a certain continuous level of agreement on a topic, like for example regarding the involvement of a country in a war, the production of nuclear energy, the choice of organic food.\\

The first model describing continuous opinion interaction is known as the Deffuant (or \emph{bounded confidence}) model \cite{deffuant2000mixing}. It has introduced the concept of bounded confidence for which people who are similar enough becomes more similar. Practically, it means that two individuals having their opinion less far than a threshold are going to have closer opinion after the interaction. Some different implementations of this model taking into account a rejection process \cite{huet2008rejection} or a different type of tolerance threshold connected to the opinion \cite{gargiulo2008can} have been proposed in the last years.\\

The interest toward opinion dynamics increases coupling these phenomena with the investigation on the topological structure of social networks. Recently this topic has been the object of many analyses, both from the theoretical and the empirical point of view (using for example the WEB 2.0 technologies).\\

Different kinds of network topologies have been tested both to prove the robustness of the opinion dynamics models and for identifying preferential channels of opinion spreading \cite{ard2004role}. At the beginning all the considered network topologies were static: namely, the connections among the persons did not vary in time. This approximation is reasonable if we consider that the processes happening on network (like opinion spreading in this case) have a different, much shorter, time scale than the process that changes the structure of network (rewiring mechanism, cutting of links).  Incidentally many works have recently been done regarding evolving network topology and their adaptation to the social background \cite{gross2008adaptive}: as people can influence each other to induce a change of mind, the difference of opinion on some very important topics can also lead to the breaking of a social contact. In other terms, since people prefer to be surrounded by persons sharing similar opinion, it is quite likely that the change of opinions due to the opinion dynamics processes can lead to the change of the network structure. An interesting analysis of the co-evolution of opinions and networks is presented in \cite{kozma2008consensus}.\\

Regarding social network analysis, sociologist and network scientists agree on the fact that social networks present sometimes community based structures: analyzing networks at different scales, it is possible to identify groups of persons much more interconnected among them than with the rest of society \cite{girvan2002community}. Many different algorithms have been created to identify communities on large networks \cite{2009community} and many models have been proposed to explain the mechanism leading to the formation of such underlying structures \cite{palla2007quantifying}.\\

In this paper we consider a community (or group) structure of individuals changing their opinions and changing groups if theirs is not sufficiently homogeneous regarding the opinion of their individuals. An individual changing groups changes at the same its social network. We will see that this coupled model lead to some interesting organizational patterns at the society, but also at the group, levels: it avoids the isolation of small extremist groups; it can also induce a very hierarchized society with very large and very small communities representing each the centered and the extreme opinions.

While the part two presents how we build the social network of each individual, the part three details the dynamical model of the individual. The next part shows and explains the results we obtained studying the model. Finally, a conclusion sums-up the model and its results. 

\section{The network setup}

Looking at the local interactions of the individuals it is possible to describe the social network in a  determined society. By the way, observing the social networks at a different scale it is usually possible to observe a  precise structure of communities where individuals have more connections among them than with the rest of society.  Many algorithm have been developed to find these community inside large networks.\\

We start our analysis from a different point of view: we build a network that already incorporates the group structures. We consider $N$ agents and a fixed number $G$ of groups - imagine, for example, the possible political parties in a nation. Each person at the beginning can choose randomly the membership in one of the $G$ groups. The network is initially constructed linking together all the agents members of the same group. Moreover each person has also the possibility of being connected with someone that is not a member of its own group. Therefore, some other links between each agent and the agents outside from its group are added with a probability $p_{ext}$. 
Since the membership inside the groups is randomly chosen, at the beginning all the groups have on average the same size ($S=N/G$).\\

Together with the network structure we are interested in studying some kind of dynamical process on the network. We focus on an opinion dynamics process and therefore we need to initialize the opinion of the agents. We use a continuous opinion framework and we attribute to each agent a random opinion in the range $\vartheta_i\in[0,1]$. Since the opinion is initialized randomly, it results that the average opinion of each group, at the beginning, is around $O_I\sim0.5$.

\section{The dynamical model}

The aim of this paper is to study simultaneously the evolution of the network topology - and therefore the properties of the group structures - and the opinion dynamics process.  The opinion of a person, indeed, has an influence for its preferences about the connections: people prefer to be linked with someone with a similar opinion (homophily). On the other side, the opinion dynamics is a contact process that take place on the links of the social network and therefore it is influenced by the topology of the network. This double feedback is realized including in the model two dynamical modules: the first one concerns opinion dynamics on the network, and the other one the local changes in the network topology. At each time step each agent can choose with an equal probability which module to perform.

\subsection{The dynamics of opinions}

To model the opinion dynamics we used an extremely known model for opinion dynamics: the Bounded confidence (BC) model by Deffuant et al \cite{deffuant2000mixing}. An agent, selected to perform opinion dynamics, decides to interact with one randomly chosen neighbour, whatever it represents a external or an internal group link. According to the BC, the two interacting agents \textit{i} and \textit{j} influence each other if their opinions differ from less than a fixed threshold $\varepsilon$. When they influence each other, their opinions become more similar:

\begin{equation}
\mbox{if }|\vartheta_i(t)-\vartheta_j(t)|<\varepsilon \qquad \left\{\begin{matrix}
 \vartheta_i(t+1)= \vartheta_i(t)+\mu(\vartheta_j(t)-\vartheta_i(t)) \\
   \vartheta_j(t+1)= \vartheta_j(t)+\mu(\vartheta_i(t)-\vartheta_j(t))
\end{matrix}\right.
\end{equation}

where $\vartheta_i$ is the opinion of agent $i$, $\vartheta_j$ the opinion of agent $j$ and $\mu$ the speed coefficient. This model has been extensively studied and all the details of the dynamics are known. The BC model presents different behaviours according to different values of $\varepsilon$; in particular four types of bifurcations separating different types of behaviours have been identified in \cite{ben2003bifurcations} and \cite{lorenz2007continuous}:   
\begin{itemize}
	\item appearance of two minor clusters symmetrically from the central one at $\varepsilon\sim0.5$ (transition from consensus to pluralism);
	\item creation of two major side clusters from the central one at $\varepsilon\sim 0.266$;
	\item separation of a minor central cluster at $\varepsilon\sim0.222$; 
	\item growth of the central cluster and shift to extremist positions of the two side clusters $\varepsilon\sim0.182$.
\end{itemize}

In particular it has been showed in \cite{fortunato2004universality} that the transition between total consensus and pluralism at $\varepsilon=0.5$ is very robust according to the network topology: it remains the same if the dynamics happens on complete graphs, lattices, random graphs and scale free networks.

\subsection{The dynamics of membership}  

The changes of the local topology of the networks are driven by the agents' decisions regarding the group membership. Following the dissonance theory of Festinger \cite{festinger1957Dissonance}, we assume that an agent with an opinion very different from the average of its own group ($|\vartheta_i-O_I|>\varepsilon$) can feel uncomfortable in this contest and can decide to change group. The choice of the new group will be done between  the set of groups on which he can retrieve some information, namely the groups in which he has some external connections.  The choice of the new group ($J$) happens with a probability:
\begin{equation}
 P_{i\rightarrow J}=\frac{1-|\vartheta_i(t)-O_J(t)|}{\sum_{J\supset j\in \mathcal{V}(i)}\left(1-|\vartheta_i(t)-O_J(t)|\right)} 
\end{equation}

where $\mathcal{V}(i)$  is the neighbourhood of node $i$.\\

After the membership has changed, all the connections of the agent $i$ are re-initialized: it is connected to all the members of the new group and, with probability $p_{ext}$ with the agents outside the new group. 

\section{Results}

In the following we will show the results of the simulations. In all the cases we used a population of \textit{N}=5000 individuals, divided into \textit{G}=500 groups: therefore, each group contains on average 1\% of the population at the initialisation time.  We fixed the probability of connection between nodes that are not members of the same group at  the value $p_{ext}=0.001$; with this setup, at the beginning, each node have on average 9 internal connections and 5 external connections.\\

The parameter $\mu$ of the bounded confidence model is fixed to the value $\mu=0.5$ as in most cases in the literature. All the simulations are repeated for 100 realizations. For each realizations, a sufficient number of steps is executed until we are sure that the model has reached its equilibrium state. The group sizes are randomly initialised following a uniform law. Samely, the opinions are uniformly distributed inside the groups.\\

The model presents interesting results, both at the global level, where we can observe substantial differences with the Deffuant BC model, and regarding the evolution of the group structures. We will present the different results in two different sections.

\subsection{BC model on adaptive networks}

\begin{figure}[h]
\epsfig{file=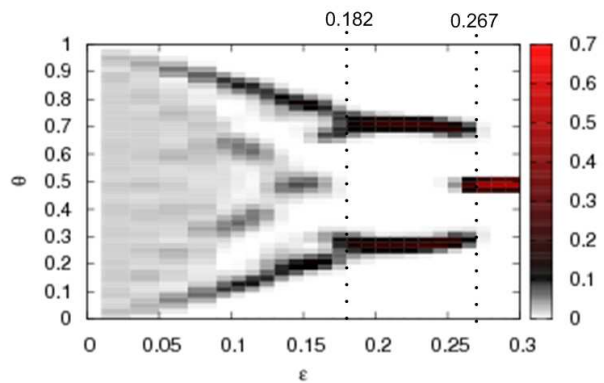, width=8.5cm}
\caption{Opinion density for different values of $\varepsilon$.}
\label{bifurcation}
\end{figure}

The Figure \ref{bifurcation} represents the density of opinion at the final state as a function of the parameter $\varepsilon$. This density has been calculated on the 100 replications for each set of parameter values.

We can observe that, when the opinion dynamics is coupled with a co-evolutive network structure the system presents a sudden transition between a state of consensus and a state of polarization of opinion (namely two major opinion clusters) at $\varepsilon\sim 0.27$. Moreover the polarized status remains until $\varepsilon=0.18$ where three macroscopic clusters (a central one plus two extremist clusters) appear.\\

Therefore the critical behaviour of the BC model on adaptive networks presents some differences with the traditional one studied by Ben Naim and Lorenz respectively in \cite{ben2003bifurcations} and \cite{lorenz2007continuous}. In fact the adaptivity of the network and the group structures avoid the formation of minor clusters: the fact that the individuals can preferentially choose neighbours (prefering persons with similar opinion) increases the probability of discussing with someone inside the tolerance bound.\\
\\

\begin{figure}[h]
\epsfig{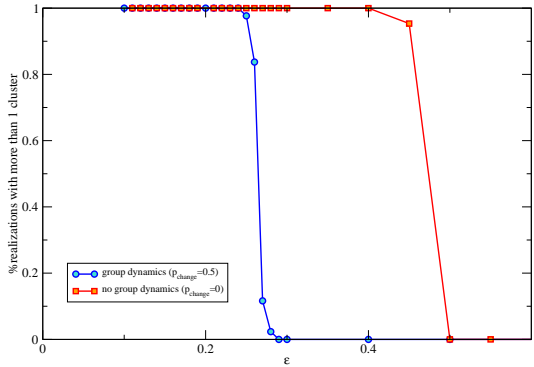}
\caption{Percentage of realizations with more than one opinion cluster for adaptive and static networks.}
\label{p_cons}
\end{figure}

The Figure \ref{p_cons} displays the percentage of realizations with more than one opinion clusters. As we can observe, in case of static networks (where group dynamics is not performed, $p_{change}=0$) such transition happens at $\varepsilon=0.5$ as it happens on all static network topologies. It is not the case for the adaptive structure, where the complete consensus is reached for $\varepsilon\sim 0.267$.\\

As explained in \cite{Gargiulo:arXiv0912.2821} exhibiting some time trajectory of a population evolving following the Deffuant opinion dynamic, and from \cite{ben2003bifurcations}, \cite{laguna2004minorities} \cite{lorenz2007continuous}, the convergence for this zone of $\varepsilon$ values happens very quickly. Some individuals remain on the border of the attitude space, especially when $\varepsilon$ is comprised between 0.267 and 0.5. They have been "forgotten" by the others due to the high speed of the dynamics which is $\mu$=0.5. These "forgotten" extremists are called "minor clusters". In the present adaptive network bounded confidence model, the convergence is slower for these specific values. Nobody is "forgotten" by the dynamics. Indeed, all people situated on the extrema of the opinion space are susceptible to change groups because they are far from the average opinion of their group. Practically the stochasticity of the model implies that the average of each group at the beginning of the simulation is around 0 and varies a bit. Then, there is always a group whose the average opinion is close enough to allow extremists to join the other people in their convergence to the centre. That is the reason why minor clusters do not exist in this "group" version of the Deffuant model.\\

It can be interesting to approach this matter from the behaviour of the Hegselman and Krause model \cite{hegselmann2002opinion} which is also a bounded confidence model with a pseudo-group approach. Indeed, the group of an individual is dynamic in this instance and corresponds to all the individuals situated at an opinion distance around the individual of almost epsilon. The individual interacting with its group adopts its average opinion. This Hegselman and Krause version of the bounded confidence model also does not exhibit some minor clusters.\\

Let's notice than the densities shown in the Figure 1 makes difficult to count the number of opinion cluster compared to the graph presented by \cite{lorenz2007continuous}. However, Lorenz used a deterministic version of the Deffuant BC model while we use an individual-based stochastic version. Using a stochastic version makes the clusters varying a bit of position in the opinion space from one realization to the other. That explains why it is difficult in our graph for low value of $\varepsilon$ to read to number of opinion clusters.\\ 

\subsection{Group hierarchy and opinion segregation}

At the moment of the initialization all the groups have homogeneous sizes and the opinions are uniformly distributed inside the groups. The system evolution exhibits interesting results, both at the level of the group size and at the level of the distribution of the opinions inside the groups.\\

\begin{figure}[h]
\epsfig{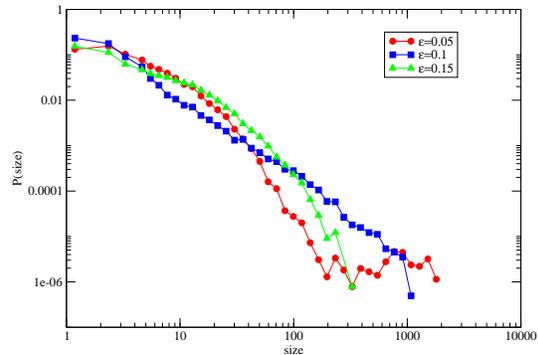}
\caption{Group size distribution for $\varepsilon=0.05,0.1,0.15$}
\label{p_size}
\end{figure}

The Figure \ref{p_size} represents the group size distribution for different values of $\varepsilon$. It is evident from the figure that in the selected cases the group sizes present a strong heterogeneity. Some macroscopic groups, containing a large part of the population, are formed with the dynamical process. Simultaneously many groups are populated by single (or extremely few) individuals. All the three cases of the Figure \ref{p_size} present a situation of hierarchy between the groups. \\
\\

\begin{figure}[h]
%\centering
\epsfig{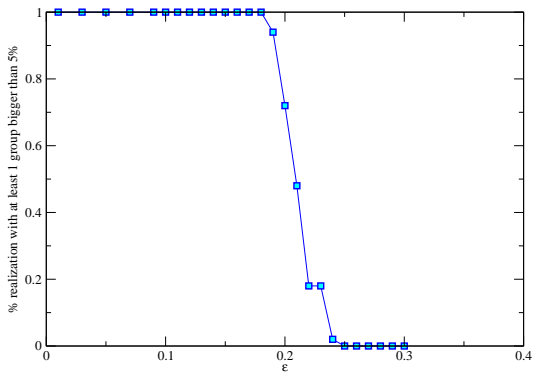}
\caption{Percentage of realizations with at least one group containing more than 5\% of the population.}
\label{perc_more5}
\end{figure}

The Figure \ref{perc_more5} shows the range of $\varepsilon$ values for which this hierarchical structure is present, giving the percentage of realisations containing at least one group representing more than 5\% of the population. Remind that the group contains about 0.2\% of the population at the initialization time. We can observe that a transition happens around $\varepsilon\sim 0.2$. After this value the groups maintain, on average, the same size as at the beginning and the group size distribution is described by a Poissonian Law with average value $N/G$.\\
\\

During a simulation, we distinguish two stages explaining the hierarchisation of the groups: a first one where people situated on the extrema tend to abandon their position in favour of the more centred opinion clusters; a second one where individuals get together in a same group. The first is the same than the one leading to the suppression of the minor clusters presented earlier in this paper. The second stage occur when the opinion clusters are formed inside the group. At this time, most of the individuals remain unsatisfied by their own group because they are far from its average opinion. Thus, as they are now unable to change their opinion, they change groups using their external links. If a group is a little larger in size than another, the individual has more external neighbours in this group. Thus the probability that this group would be chosen increases with the size of the group, and at the same time it increases with the closeness of its average opinion. As the individual does not change opinion clusters when it changes groups, each time it chooses a group, it contributes making the average opinion of the group closer to the one of the opinion cluster. It also contributes increasing the size of the group and then, the numbers of external links the other individuals are going to have to this group. That is the beginning of a recursive phenomenon leading to the existence of few very large opinion clusters (sometimes only one) positioned in only a few groups (one if there is only one large opinion cluster).\\

\begin{figure}[h]
\epsfig{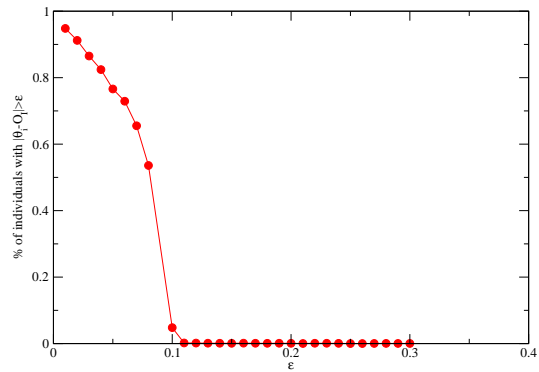}
\caption{Percentage of individuals susceptible to change group.}
\label{perc_cohesive}
\end{figure}

The Figure \ref{perc_cohesive} gives us indication about how the opinions of the population of each group evolve from the initially uniform situation. It represents the number of individuals that are not compatible with their own group, namely the individuals susceptible to change groups due to their opinion distance with the average one of their group ($|\vartheta_i-O_I|>\varepsilon$). Notice that, even if these agents are susceptible to change groups, they are able to do this only if another group has an average opinion closer to theirs. By the way this measure is a good indicator to express the opinion homogeneity inside the groups. Indeed, at the end of the simulation, opinions are divided into clusters distant among them from at least $\varepsilon$. Thus, if there is more than one opinion cluster inside the group, some agents are automatically susceptible to change due to their high distance to the average opinion of the group.\\

Figure \ref{perc_cohesive} shows that at $\varepsilon=0.11$ the situation of opinion homogeneity inside the groups is reached: namely all the groups contain only persons with an opinion very close to the average. The opinions are segregated into different groups (each group contains just an opinion even if at global level a larger variety of opinions is present). This is therefore the transition to total consensus inside each group.\\

Coming back to the Figure \ref{perc_more5} to compare it to the Figure \ref{perc_cohesive}, one can see that the hierarchization of the group regarding their size is a very stable phenomenon from $\varepsilon=0.11$ to $\varepsilon=0.2$. Indeed, for this range of value, the individuals don't change neither groups, nor opinion cluster. The situation seems different from the $\varepsilon$ values lower than 0.11. However, this zone is not studied here even if it deserves more investigations.

\section{Conclusion}

We have proposed a model coupling the classical opinion dynamics of the bounded confidence model proposed by Deffuant with an adaptive network forming a community or group structure. At each step, an individual can decide if it changes groups or interact on its opinion with one of its internal or external neighbour. If it decides to look at the group level, it changes groups if its opinion is far from the average of its group from more than $\varepsilon$. If it is the case, it joins the group which has proportionnally the closest average opinion from its. If it decides to interact with one of its neighbour, it becomes closer in opinion to it when its opinion and the one of the selected-to-interact neighbour are less distant from $\varepsilon$.\\

From the study of this coupled model, we discover some surprising behaviours compared to the known behaviour of the Deffuant bounded confidence model(BC):

\begin{itemize}
	\item The coupled model exhibits a total consensus for an $\varepsilon$ value lower than the BC model. That is linked to the capacity of the coupled model to suppress the minor clusters positionned in the BC model on the extrema of the opinion space. In social psychology, groups are known as a source of cohesion and avoidance of the isolation. Thus, that is a very interesting fact that the introduction of groups in the BC model suppresses the isolated individuals. 
	\item The distribution of sizes of the groups changes: some groups become larger while other decrease in size, sometimes until containing only one individual. This can be mainly explained by the fact that people unsatisfied by their group have a preferential external attachment to the larger group. Thus, when a group is a little bit larger due to the stochasticity of the model, it increases its probability to welcome unsatisfied people. More individuals come in, more larger it is and more probable the new arrivals in are. That sounds quite realistic. Indeed, a lot of people tend to change groups when they are in a dissonant situation and to join a larger group which appears more comfortable when it exists. 
	\item From the point of vue of the groups, the consensus remains for a large set of $\varepsilon$ values while, looking at the population level, there are a lot of opinion clusters. Then, each group does not only correspond to a subpopulation exhibiting the same behaviour than the whole population. In politics, we often see that a given opinion about an issue is the attribute of a given group. At the global level, each opinion present in the population correspond to a group's one. That tends to be not the truth when the opinion related issue is very important for people. In this case, the group level is less important for people and they prefer changing groups and remaining in the same opinion community.
\end{itemize}
  
These first results are interesting. However, not all the parameters of the model have been studied. That is now necessary to investigate more the model. It would be especially interesting to understand what occur for low values of $\varepsilon$.

Another very important way of investigation would be to consider a different way for the individual to take into account a group membership via its social network. A work as the one done by \cite{groeber2009} can be a source of inspiration. Using the same principle to couple the BC model to a social network implying a community structure, he obtains results quite different from us. Thus, a more detailed comparison among this both work would give some knowledge about the modelling question of these processes. 

\bibliographystyle{plain}	
\bibliography{GO_refs}
\end{document}